\documentclass[12pt]{article}
\usepackage{amssymb}

\begin{document}

\newcommand{\sect}[1]{\setcounter{equation}{0}\section{#1}}
\renewcommand{\theequation}{\thesection.\arabic{equation}}
\newcommand{\be}{\begin{equation}}
\newcommand{\ee}{\end{equation}}
\newcommand{\bea}{\begin{eqnarray}}
\newcommand{\eea}{\end{eqnarray}}
\newcommand{\nonu}{\nonumber\\}
\newcommand{\beano}{\begin{eqnarray*}}
\newcommand{\eeano}{\end{eqnarray*}}
\newcommand{\eps}{\epsilon}
\newcommand{\om}{\omega}
\newcommand{\vph}{\varphi}
\newcommand{\sig}{\sigma}
\newcommand{\CC}{\mbox{${\mathbb C}$}}
\newcommand{\RR}{\mbox{${\mathbb R}$}}
\newcommand{\QQ}{\mbox{${\mathbb Q}$}}
\newcommand{\ZZ}{\mbox{${\mathbb Z}$}}
\newcommand{\NN}{\mbox{${\mathbb N}$}}
\newcommand{\1}{\mbox{\hspace{.0em}1\hspace{-.24em}I}}
\newcommand{\II}{\mbox{${\mathbb I}$}}
\newcommand{\prt}{\partial}
\newcommand{\und}[1]{\underline{#1}}
\newcommand{\wh}[1]{\widehat{#1}}
\newcommand{\wt}[1]{\widetilde{#1}}
\newcommand{\mb}[1]{\ \mbox{\ #1\ }\ }
\newcommand{\half}{\frac{1}{2}}
\newcommand{\noin}{\not\!\in}
\newcommand{\rhotimes}{\mbox{\raisebox{-1.2ex}{$\stackrel{\displaystyle\otimes}
{\mbox{\scriptsize{$\rho$}}}$}}}
\newcommand{\bin}[2]{{\left( {#1 \atop #2} \right)}}
\newcommand{\alg}{{\cal A}_R}
\newcommand{\balg}{{\cal B}_R}
\newcommand{\salg}{{\cal S}_R}
\newcommand{\cf}{{\cal F}}
\newcommand{\R}{{\bf R}}
\newcommand{\hl}{{\RR}_+}
\newcommand{\C}{{\bf C}}
\newcommand{\Hil}{{\cal H}}
\newcommand{\D}{{\cal D}}
\newcommand{\brep}{{\cal F}_B}
\newcommand{\form}{\langle \, \cdot \, , \, \cdot \, \rangle }
\newcommand{\e}{{\rm e}}

\newcommand{\baral}{\bar{\alpha}}
\newcommand{\barbt}{\bar{\beta}}

\newcommand{\EE}{\mbox{${\mathbb E}$}}
\newcommand{\JJ}{\mbox{${\mathbb J}$}}
\newcommand{\MM}{\mbox{${\mathbb M}$}}
\newcommand{\ct}{{\cal T}}

\newtheorem{theo}{Theorem}[section]
\newtheorem{coro}[theo]{Corollary}
\newtheorem{prop}[theo]{Property}
\newtheorem{defi}[theo]{Definition}
\newtheorem{conj}[theo]{Conjecture}
\newtheorem{lem}[theo]{Lemma}
\newcommand{\prf}{\underline{Proof:}\ }
\newcommand{\finprf}{\null \hfill {\rule{5pt}{5pt}}\\[2.1ex]\indent}

\pagestyle{empty}
\rightline{April 2001}

\vfill

\begin{center}
{\Large\bf Spontaneous symmetry breaking in the\\[1.2ex]
  $gl(N)$-NLS hierarchy on the half line}
\\[2.1em]

{\large
M. Mintchev$^{a}$\footnote{mintchev@df.unipi.it},
E. Ragoucy$^{b}$\footnote{ragoucy@lapp.in2p3.fr}
and P. Sorba$^{b}$\footnote{sorba@lapp.in2p3.fr}}\\
\end{center}

\null

\noindent
{\it $^a$ INFN, Dipart. di Fisica dell'Univ. di
    Pisa, Via Buonarroti 2, 56127 Pisa, Italy\\[2.1ex]
$^b$ LAPTH, Chemin de Bellevue, BP 110, F-74941 Annecy-le-Vieux
    cedex, France}
\vfill

\begin{abstract}
We describe an algebraic framework for studying the symmetry
properties of integrable quantum systems on the half line.
The approach is based on the introduction of boundary
operators. It turns out that these operators both encode the boundary
conditions and generate integrals of motion. We use this
direct relationship between boundary conditions and
symmetry content to establish the spontaneous breakdown
of some internal symmetries, due to the boundary.

\end{abstract}

\vfill
\rightline{LAPTH-844/2001}
\rightline{IFUP-TH 15/2001}
\rightline{\tt hep-th/0104079}
\newpage
\pagestyle{plain}
\setcounter{page}{1}

\sect{Introduction}

\null

This paper describes a framework for studying the symmetry properties of
integrable quantum systems on the half line. As it can be expected on general
grounds, the presence of a boundary in this case has a strong impact on the
dynamics and the symmetry content of the systems. It gives rise to
a variety of boundary-related phenomena with direct applications to
impurity problems in condensed matter, dissipative quantum mechanics,
open string theory and brane physics.
The recent efforts to gain a deeper insight in these phenomena stimulated
a series of investigations \cite{GS}-\cite{AOS} in the subject. Among
others, one should mention the attempts to develop an algebraic
approach. The new strategy there is to introduce \cite{GS,FK}
the so called {\sl boundary operators}, which translate in
algebraic terms the associated quantum boundary value problem. It is
far from being a
surprise that if possible, an algebraic treatment of the boundary value problem
turns out to be simpler then
the standard  analytic one. In the context of integrable systems, the
algebraic framework provides a
further advantage - the search for general integrability preserving
boundary conditions
and their implementation become straightforward.

The concept of boundary algebra, we will be dealing below,
has been introduced in \cite{LMZ} and is
inspired by Cherednik's scattering theory \cite{C} for integrable models
on the half line. An essential feature of this algebra, denoted in
what follows by
$\balg $, is that it enables
one to  reconstruct \cite{LMZ} not only the scattering matrix, but captures
also the off-shell dynamics (correlation functions) of the system
\cite{GLM1,GLM2}.
Therefore, one can apply  $\balg$ in the study of symmetries as well.
This observation is the basic starting point of our investigation.
The most interesting property
emerging from it is that, besides encoding the boundary conditions,
the boundary operators generate also integrals of motion.
The main goal of the present work is to investigate this remarkable
feature and to explore the consequences from it.

The paper is organized as follows. In the next section we summarize
those basic features
of the boundary algebra $\balg $ and its Fock representations, which are needed
for our analysis. The general structure involved in the discussion
is illustrated by means of the $gl(N)$-invariant nonlinear Schr\"odinger
model on the half line. The quantization of the model in terms of $\balg$ is
briefly described in Section \ref{sect-BNLS}. Section
\ref{sect-NLSsym} is devoted to a purely algebraic
analysis of the symmetry content of the NLS model. In Section
\ref{sect-boundcond} we derive
a large class of boundary  conditions which preserve integrability.
The phenomenon of spontaneous symmetry breaking is established in Section
\ref{sect-symbreak}. Section \ref{n=2} contains a detailed study
of the case $N=2$. Finally,
Section \ref{sect-concl} collects our conclusions.

\sect{Boundary algebras\label{sect-Balg}}

We recall first the basic structure of a boundary algebra, referring for
details to \cite{LMZ}. $\balg $ is
an associative algebra with identity element $\bf 1$ and two
types of generators,
\be
\{a_i (k),\, a^{\ast i } (k)\, \, : \, \, i = 1,...,N,
\, \, k \in \RR \}
\label{a}
\ee
and
\be
\{b_i^j (k) \, \,  : \, \, i , j = 1,...,N, \, \, k
\in \RR \} \, .
\label{b}
\ee
called bulk and boundary generators respectively. $N$ is the number of internal
degrees of freedom, whereas $k$ stands for the momentum in the non
relativistic case
or the rapidity in the relativistic one.
(\ref{a}) and (\ref{b}) are subject to the following constraints:

\noindent (i) bulk exchange relations
\bea
a_{i_1 }(k_1) \, \, a_{i_2 }(k_2) \, \; - \; \,
     R_{i_2 i_1 }^{j_1 j_2 }
     (k_2 , k_1)\,\, a_{j_2 }(k_2)\, a_{j_1 }(k_1) & = & 0
     \, , \label{aa}\\
a^{\ast i_1 } (k_1)\, a^{\ast i_2 } (k_2) -
     a^{\ast j_2 } (k_2)\, a^{\ast j_1 } (k_1)\,
     R_{j_2 j_1 }^{i_1 i_2 }(k_2 , k_1) & = & 0
     \, , \label{a*a*} \\
a_{i_1 }(k_1)\, a^{\ast i_2 } (k_2) \; - \;
     a^{\ast j_2 }(k_2)\,
     R_{i_1 j_2 }^{i_2 j_1 }(k_1 , k_2)\,
     a_{j_1 }(k_1) & = &  \nonumber\\
     2\pi\, \delta (k_1 - k_2)\,
     \delta_{i_1 }^{i_2 }\, {\bf 1} +
     2\pi\, \delta (k_1 + k_2)\,  b_{i_1 }^{i_2 }(k_1)
     \, ; \label{aa*}
\eea
(ii) boundary exchange relations
\bea
R_{i_1 i_2}^{l_2 l_1}(k_1 , k_2)\,
b_{l_1}^{m_1}(k_1)\,
R_{l_2 m_1}^{j_1 m_2}(k_2 , -k_1)\,
b_{m_2}^{j_2}(k_2) = \nonumber\\
b_{i_2}^{l_2}(k_2)\,
R_{i_1 l_2}^{m_2 m_1}(k_1 , -k_2)\,
b^{l_1}_{m_1}(k_1)\,
R_{m_2 l_1}^{j_1 j_2}(-k_2 , -k_1)
\, ; \label{bb}
\eea
(iii) mixed exchange relations
\be
a_{i_1}(k_1)\, b_{i_2}^{j_2}(k_2) =
R_{i_2 i_1}^{l_1 l_2}(k_2 , k_1)\,
b_{l_2}^{m_2}(k_2)\,
R_{l_1 m_2}^{j_2 m_1}(k_1 , -k_2)\, a_{m_1}(k_1)
\, , \label{ab}
\ee
\be
b_{i_1}^{j_1}(k_1)\, a^{\ast i_2}(k_2) =
a^{\ast m_2}(k_2)\,
R_{i_1 m_2}^{l_2 m_1}(k_1 , k_2)\,
b_{m_1}^{l_1}(k_1)\,
R_{l_2 l_1}^{j_1 i_2}(k_2 , -k_1)\,
\, . \label{ba*}
\ee
Here and in what follows the summation over repeated upper
and lower indices is always understood. Finally, using already
standard notations, the exchange factor $R$ obeys
\be
R_{12}(k_1,k_2)\, R_{12}(k_2,k_1) = 1 \, , \label{RR}
\ee
and the spectral quantum Yang-Baxter equation (in its braid form)
\be
R_{12}(k_1,k_2) R_{23}(k_1,k_3) R_{12} (k_2,k_3)
= R_{23}(k_2,k_3) R_{12}(k_1,k_3) R_{23}(k_1,k_2)  \, .
\label{qyb}
\ee
These compatibility conditions on $R$,
together with eqs.(\ref{aa}-\ref{ba*}) define the algebra $\balg $.
Let us observe also that the generators $\{b_i^j (k)\}$ close
a subalgebra $\salg \subset \balg$,  which has been introduced by Sklyanin
\cite{S} and which will play a distinguished role in our discussion below.

It is worth mentioning that if one formally takes $b_i^j (k)\to 0$,
the relations (\ref{bb}-\ref{ba*})
trivialize, while (\ref{aa}-\ref{aa*}) reproduce the defining relations
of a Zamolodchikov-Faddeev algebra \cite{ZZ,F}, which is known to describe
factorized scattering of 1+1 dimensional integrable systems on the whole line.
The situation changes on the half line, where in order to
reproduce Cherednik's scattering amplitudes one needs
\cite{LMZ} a {\sl reflection algebra},
i.e. a $\balg $ with the additional constraint
\be
b_i^l (k)\, b_l^j (-k) = \delta_i^j \, ,
\label{rc}
\ee
which obviously prevents the vanishing of all boundary operators. The condition
(\ref{rc}) has important consequences; it implies that the mapping
\be
\varrho \, :\, a_i (k) \mapsto b_i^j (k) a_j (-k )
\, , \label{r1}
\ee
\be
\varrho \, :\, a^{\ast i }(k) \mapsto a^{\ast j } (-k )
b^i_j (-k)
\, , \label{r2}
\ee
\be
\varrho \, :\, b_i^j (k) \mapsto b^j_i (k)
\, , \label{r3}
\ee
leaves invariant the constraints (\ref{aa}-\ref{ba*}) and extends therefore
to an automorphism on $\balg $.
$\varrho $, which is called in what follows {\sl reflection automorphism},
has a direct physical
interpretation in scattering theory: it provides a mathematical
description of the intuitive physical picture that bouncing back
from a wall, particles change the sign of their momenta (rapidities). In
fact, the two elements $a^{\ast i }(-k)$ and $a^{\ast j }
(k) b^i_j (k) $ are $\varrho $-equivalent,
\be
a^{\ast
i }(-k) \sim a^{\ast j } (k) b^i_j (k) \, . \nonumber
\ee
Combined with (\ref{bb}), eq.(\ref{rc}) turns out to be essential
also in the description of symmetries.

{}For constructing representations of $\balg $, it is essential to
recognize an involution on it. Let us consider for this purpose
the mapping $I$ defined by
\be
I \, : \,  a^{\ast i }(k) \mapsto a_i (k) \, , \qquad
a_i (k) \mapsto a^{\ast i }(k)\, , \qquad
b_i^j  (k) \mapsto b_j^i (-k)\, . \label{i}
\ee
When extended as an antilinear antihomomorphism on $\balg $,
$I$ defines an involution, provided that $R$ satisfies
\be
R^{\dagger j_1 j_2}_{\, \, i_1 i_2}(k_1 , k_2 ) =
R^{j_1 j_2}_{i_1 i_2}(k_2 , k_1 )
\, . \label{Ha}
\ee
Here and in what follows the dagger stands for matrix Hermitian
conjugation, i.e.
\be
R^{\dagger j_1 j_2 }_{\, \, i_1 i_2} (k_1,k_2)
\equiv {\overline R_{j_1 j_2 }^{i_1 i_2}} (k_1,k_2)
\, ,
\ee
the bar indicating complex conjugation. The condition (\ref{Ha}) is
known in the literature on factorized scattering as {\sl Hermitian
analyticity}.
We observe in passing that a larger class of involutions with the appropriate
generalization of (\ref{Ha}) have been introduced in \cite{LMZ}.

Since $\balg$ is an infinite algebra, for the moment our considerations
are a bit formal. In order to give them a precise mathematical meaning,
one can construct \cite{LMZ} a class of Fock representations of the
reflection algebra with involution $\{\balg , I\}$,
characterized by the following requirements:
\medskip

\noindent 1. The representation space $\Hil$ is a Hilbert space
with scalar product $\form $;
\medskip

\noindent 2. The generators (\ref{a},\ref{b}) are operator valued distributions
with common and invariant dense domain $\D\subset \Hil$, where
eqs.(\ref{aa}-\ref{ba*}) hold.
The involution $I$ defined by eq.(\ref{i}) is realized as a
conjugation with respect to $\form $;
\medskip

\noindent 3. There exists a vector (vacuum state) $\Omega \in \D$
which is annihilated by $a_i (k)$. $\Omega $ is
cyclic with respect to $\{a^{\ast i } (k)\}$ and
$\langle \Omega\, ,\, \Omega \rangle\, = 1$.
\medskip

These requirements imply that the c-number distributions
\be
B_i^j (k) \equiv \langle
\Omega \, ,\, b_i^j (k)\Omega \rangle
\, , \label{B}
\ee
satisfy
\be
B_{\, \, \, i}^{\dagger j }(k) =
B_i^j (-k)\, , \label{bHa}
\ee
which, being an analog of (\ref{Ha}), is called {\it boundary}
Hermitian analyticity.
One can show moreover, that the vacuum vector
$\Omega $ is unique (up to a phase factor) and satisfies
\be
b_i^j (k)\, \Omega =  B_i^j (k)\, \Omega
\, . \label{bO}
\ee
Therefore, taking the vacuum expectation value of
eqs.(\ref{bb},\ref{rc}), one finds
\bea
&R_{i_1 i_2}^{l_2 l_1}(k_1 , k_2)\,
    B_{l_1}^{m_1}(k_1)\,
    R_{l_2 m_1}^{j_1 m_2}(k_2 , -k_1)\,
    B_{m_2}^{j_2}(k_2) = \nonumber \\
&B_{i_2}^{l_2}(k_2)\,
    R_{i_1 l_2}^{m_2 m_1}(k_1 , -k_2)\,
    B^{l_1}_{m_1}(k_1)\,
    R_{m_2 l_1}^{j_1 j_2}(-k_2 , -k_1)
    \, , \label{RBRB}
\eea
\be
B_i^l (k)\, B_l^j (-k) = \delta_i^j \, .
\label{rcB}
\ee
We thus recover at the level of Fock representation the
{\sl boundary} Yang-Baxter equation (\ref{RBRB}), originally derived
in \cite{C}. Because of
(\ref{rcB}), we will refer to $B$ as {\sl reflection matrix}.

Given a reflection algebra $\{\balg , I\}$, its Fock representations are
classified by all possible reflection matrices $B$, which actually parametrize
the boundary conditions. More precisely, any $B$-matrix satisfying
eqs.(\ref{bHa},\ref{RBRB},\ref{rcB}), defines a Fock representation $\brep $,
whose explicit construction is given in \cite{LMZ}. Each $\brep $ uniquely
defines in turn a unitary scattering operator $S$, corresponding to
the integrable
model described by $\{\balg , I\}$. The emerging picture is therefore
the following.
The mere fact that we are dealing with an integrable system with
boundary shows up
at the algebraic level, turning the Zamolodchikov-Faddeev algebra into
reflection algebra. The details of the boundary condition enter
at the representation level through the reflection matrix $B$. We stress
at this point another sharp difference between reflection and
Zamolodchikov-Faddeev algebras, the latter admitting a unique
(up to unitary equivalence) Fock representation.

Let us turn finally to the question of symmetries, considering for instance
the simplest non relativistic Hamiltonian coming in mind, namely
\be
H = \int_{-\infty}^\infty dp\, p^2 a^{*i}(p) a_i (p) \, . \label{H}
\ee
Using eqs.(\ref{ab},\ref{ba*}), one easily verifies that
\be
\left [H\, ,\, b_i^j (k)\right ] = 0 \, . \label{com}
\ee
Therefore, besides capturing the presence of boundaries, $b_i^j (k)$
generate integrals of motion, whose algebra is encoded in the boundary exchange
relations (\ref{bb}).

It will be our main objective
in what follows to examine thoroughly this observation on both purely algebraic
(section \ref{sect-NLSsym}) and Fock representation (section
\ref{sect-boundcond}) levels. In order to
gain some intuition
from a concrete example, we find it useful to illustrate first the
above abstract setup
by means of the $gl(N)$-invariant nonlinear Schr\"odinger
($gl(N)$-NLS) model on the
half line.

Note also that the property (\ref{com}) can be extended to a wide
class of operators $H_{n}$:
\be
[H_{n},\, b_i^j (k) ] = 0 \mb{with} H_{n} = \int_{-\infty}^\infty
dp\, p^n a^{*i}(p) a_i (p)
\ee
leading to the notion of hierarchy. However, we want to stress that
one has to check the vanishing of the commutators between different
$H_{n}$ to get a true hierarchy. A direct calculation shows:
\be
[H_{n},H_{m}]=\left(\rule{0ex}{1.2ex}(-1)^n-(-1)^m\right)\,
\int dp\, p^{m+n} a^{*i}(-p)
b_i^j (-p) a_j (p)
\ee
Thus, it is only the $H_{n}$ of same parity that commute one with each
other, and the hierarchy should be associated to the "even"
Hamiltonians $H_{2n}$. We reject the hierarchy associated to the
"odd" Hamiltonians $H_{2n+1}$, because they satisfy
\be
H_{2n+1}=0\mb{in the Fock representation}
\ee
In what follows, we will concentrate on the Hamiltonian $H_{2}$, but
one has to keep in mind that the properties will apply to the whole
hierarchy $\{H_{2n},\ n\in\NN\}$.

\sect{The NLS model on the half line\label{sect-BNLS}}

The dynamics of the $gl(N)$-NLS model can be described by
a $N$-component field $\Psi_i (t,x)$, satisfying
\be
(i\partial_t + \partial_x^2 )\Psi_i (t,x) =
2g\, \Psi^{\dagger j} (t,x) \Psi_j (t,x) \Psi_i (t,x) \;
, \; \; \; g > 0 \, ,
\label{eqm}
\ee
on the half line $\hl = \{x\in \RR\, :\, x>0 \}$.
One must fix in addition the boundary conditions. We start by requiring
\be
\lim_{x \downarrow 0} ( \partial_x - \eta ) \Psi_i (t,x) = 0 \; , \;
\; \;  \eta \geq 0\; ; \label{bc0}
\ee
\be
\lim_{x \to \infty } \Psi_i (t,x) = 0 \; .
\label{bcinf}
\ee
Eq. (\ref{bc0}) is the so called mixed boundary condition; for $\eta
= 0$ and in
the limit $\eta \rightarrow \infty $ it reproduces the familiar
Neumann and Dirichlet conditions respectively. We will focus in this section
on the boundary value problem (\ref{eqm}-\ref{bcinf}),
postponing the consideration of more general integrability preserving
boundary conditions to section \ref{sect-NLSsym}.

In spite of the fact that the NLS model is among the most studied
integrable systems,
to our knowledge there exist only few papers addressing the problem on $\hl $.
References \cite{S} and \cite{Fo} deal essentially with the proof of
integrability,
whereas \cite{GLM1} and \cite{GLM2} are concerned with the
construction of the quantum field $\Psi $ and the relative off-shell
correlation functions. We will follow below \cite{GLM1, GLM2},
recalling first the basic structures which enter the second quantization
of the boundary value problem (\ref{eqm}-\ref{bcinf}):

\medskip

\noindent 1. A Hilbert space $\{\Hil_{g,\eta},\, \form \}$
describing the states of the system;

\noindent 2. An operator valued distribution $\Psi(t,x)$ which
acts on an appropriate dense domain $\D\subset \Hil_{g,\eta}$ and
satisfies:

(a) the canonical commutation  relations
\be
[\Psi_i (t,x)\, ,\, \Psi_j (t,y)]=
[\Psi^{*i}(t,x)\, ,\, \Psi^{* j}(t,y)]= 0 \, ,
\label{cac1}
\ee
\be
[\Psi_i (t,x)\, ,\, \Psi^{* j}(t,y)] = \delta_i^j
\, \delta (x-y)
\, , \label{cac2}
\ee
where $\Psi^*$ is the $\form $-Hermitian conjugate of $\Psi $;

(b) the equation of motion
\be
(i\partial_t + \partial_x^2 )\langle \varphi \, ,\, \Psi_i (t,x)
\psi \rangle
= 2g\, \langle \varphi \, ,\, :\Psi_j \Psi^{*j} \Psi_i :
(t,x) \psi \rangle \, ,
\ \  \forall \; \varphi , \psi \in \D \, ,
\label{qeqm}
\ee
where $:\, \dots \, :$ denotes a suitably defined normal product;

(c) the boundary conditions
\be
\lim_{x \downarrow 0} \left ( \partial_x - \eta \right )
\langle \varphi \, ,\, \Psi_i (t, x) \psi \rangle = 0\, , \qquad
\forall \; \varphi , \psi \in \D \, ,
\label{mbc0}
\ee
\be
\lim_{x\to\infty }
\langle \varphi \, ,\, \Psi_i (t, x) \psi \rangle = 0 \, , \qquad
\forall \; \varphi , \psi \in \D \, ;
\label{mbcinf}
\ee

\noindent 3. A fundamental state $\Omega \in \D$, which is cyclic
with respect to $\Psi^*$.
\medskip

It has been shown in \cite{GLM2} that all these structures can be explicitly
realized in terms of the reflection algebra $\balg $ defined by the
exchange matrix
\be
R_{i_2 i_1}^{j_1 j_2}(k_1 , k_2) =
\frac{1}{k_1-k_2 +ig} \left[ig\, \delta_{i_1}^{j_1}\,
\delta_{i_2}^{j_2}
+ (k_1-k_2)\, \delta_{i_1}^{j_2}\,
\delta_{i_2}^{j_1}\right] \, ,
\label{SchR}
\ee
and the associated Fock representation $\brep $ with reflection matrix
\be
B_i^j (k) = \frac{k-i\eta}{k+i\eta}\, \delta_i^j \, .
\label{SchB}
\ee
One has to identify for this purpose $\Hil_{g,\eta}$ with
the Fock space, $\Omega $ with the Fock vacuum and $\form $ with the
scalar product of
the representation $\brep $. Moreover, the quantum field $\Psi $
admits the expansion
\be
\Psi_i (t,x) = \sum_{n=0}^{\infty} (-g)^n \Psi_i^{(n)}(t,x) \, ,
\label{exp}
\ee
where
\bea
\Psi_i^{(n)}(t,x) =
\int_{-\infty}^\infty \prod_{i=1\atop j=0}^n {dp_i \over {2\pi}}{dq_j
\over {2\pi}}
{ {\rm e}^{ i\sum_{j=0}^n (xq_j-tq^2_j) -i\sum_{i=1}^n (xp_i-tp^2_i) }
\over \prod_{i=1}^n \left[ (p_i - q_{i-1} - i\epsilon )\,
(p_i - q_{i} - i\epsilon ) \right]}\cdot \nonumber \\
a^{*j_1}(p_1)\cdots a^{*j_n}(p_n) \, a_{j_n}(q_n)\cdots
a_{j_1}(q_1)
a_i(q_0)
\; .
\label{nth}
\eea
$a_i$ and $a^{*j}$ representing the $\balg$-generators in $\brep$.
By construction, the domain $\D$ involves only vectors with finite,
although arbitrary large, particle number.
Combining this property with the normal ordered form of $\Psi^{(n)}$,
one concludes
that the series (\ref{exp}) converges in mean value on $\D$.

Eq. (\ref{nth}) is the quantum inverse scattering transform for the
$gl(N)$-NLS model
on $\hl$. The field $\Psi$ admits (strong) asymptotic limits, giving
raise to the following
asymptotic states:
\be
|k_1,i_1;...;k_n,i_n \rangle^{\rm in} =
a^{*i_1}(k_1)...a^{*i_n}(k_n)\Omega \; ,
\quad  k_1<...<k_n<0 \; ,
\label{in}
\ee
\be
|p_1,j_1;...;p_m,j_m\rangle^{\rm out} =
a^{*j_1}(p_1)...a^{*j_m}(p_m)\Omega \; ,
\quad  p_1>...>p_n>0 \; .
\label{out}
\ee
The vectors (\ref{in}) and (\ref{out}) generate the
asymptotic in- and out-spaces respectively. Both of these spaces are dense
in $\Hil_{g,\eta}$ and the model is asymptotically complete in the range
$g>0$ and $\eta\geq 0$. The connection with Cherednik's scattering
theory \cite{C}
is obtained on the level of scattering amplitudes
\be
{}^{\rm out} \langle p_1,j_1,...,p_m,j_m
|k_1,i_1,...,k_n,i_n \rangle ^{\rm in} \; .
\label{ampl}
\ee
The explicit form of (\ref{ampl}) is easily derived and involves a
product of $R$ and $B$-factors: any $R$-factor
describes a two-body scattering in the bulk $\hl$, while the $B$-factors
take into account the reflection from the boundary.

It is worth stressing that the time evolution of the field $\Psi(t,x)$ is
generated precisely by the Hamiltonian (\ref{H}). In fact, by means
of eq. (\ref{nth})
one gets
\be
\Psi_i (t,x) = \e^{itH}\, \Psi_i (0,x)\, \e^{-itH} \, .
\label{tev}
\ee
Therefore, according to our discussion at the end of the previous section,
the boundary operators $b_i^j (k)$ generate
integrals of motion for the $gl(N)$-NLS model on $\hl$. The relative
algebra, which follows from eq. (\ref{bb}) by inserting
the specific exchange factor (\ref{SchR}), will
be investigated in the next section.

In conclusion, we point out that the quantum field (\ref{exp},\ref{nth})
solves (\ref{qeqm}) for any reflection matrix $B$ satisfying eq.(\ref{RBRB})
with $R$ given by (\ref{SchR}). One can also demonstrate \cite{GLM2} that
for any $\varphi ,\, \psi \in \D$, there exist $N$ square integrable
functions $\chi_i (k)$, such that
\be
\langle \varphi \, ,\, \Psi_i (t, x) \psi \rangle =
\int_{-\infty}^\infty \frac{dk}{2\pi} \, e^{-itk^2}\, \left
[\delta_i^j \e^{ixk} +
B_i^j (k) \e^{-ixk} \right ] \, \chi_j (k) \, .
\label{gbc}
\ee
This identity implies not only (\ref{mbc0}) and (\ref{mbcinf}), but
allows to analyze
(see section 5) the boundary condition at $x=0$ for general $B$-factors.

\sect{The NLS symmetry algebra\label{sect-NLSsym}}

Now we bring our attention to the $\salg$ subalgebra defined from the boundary
generators (\ref{b}) and exchange relation (\ref{bb}), when the R-matrix is
chosen as in (\ref{SchR}). In the standard notations, the R-matrix takes
the form:
\be
R_{{12}}(k)=\frac{1}{k+i\, g}\left( k\, \II\otimes \II+ig \,P_{{12}}\right)
\label{truc}
\ee
where $\II$ is the $N\times N$ identity matrix,
$P_{{12}}\equiv P= \sum_{i,j=1}^N \, E_{ij}\otimes E_{ji}$ is the flip
operator, and $E_{ij}$ is the $N\times N$ matrix with 1 at position
$(i,j)$.

\subsection{Calculation of the $\salg$ algebra}
Let us rewrite (\ref{bb}) as:
\be
R(k_{1}-k_{2})\, b_{1}(k_{1})\, R(k_{1}+k_{2})\, b_{2}(k_{2})\, =
b_{2}(k_{2})\, R(k_{1}+k_{2})\, b_{1}(k_{1})\, R(k_{1}-k_{2})\,
\label{rbrb}
\ee
or as commutation relation
\bea
[b_{1}(k_{1}),b_{2}(k_{2})] &=&\frac{ig}{k_{1}-k_{2}}\left(\rule{0ex}{1.2em}
b_{2}(k_{2})b_{1}(k_{1})P-Pb_{1}(k_{1})b_{2}(k_{2})\right)+\nonu
&&+\frac{ig}{k_{1}+k_{2}}\left(\rule{0ex}{1.2em}
b_{2}(k_{2})Pb_{1}(k_{1})-b_{1}(k_{1})Pb_{2}(k_{2})\right)+\nonu
&&-\frac{g^2}{k^2_{1}-k^2_{2}}\ \left[\rule{0ex}{1.2em}
b_{2}(k_{2}),b_{2}(k_{1})\right]
\label{com.bubv}
\eea
where
\be
b_{1}(k)=\sum_{i,j=1}^N b_{ij}(k)\, E_{ij}\otimes\II
\mb{and} b_{2}(k)=\sum_{i,j=1}^N b_{ij}(k)\, \II\otimes E_{ij}
\label{eq:bseries}
\ee
Note that it enters in the algebras of type $ABCD$, introduced in
\cite{FM}, and as such is well-defined, since it obeys the
corresponding consistency conditions (see \cite{FM} for details).

We develop $b(k)$ as a formal power series in $k^{-1}$:
\be
b(k)=\sum_{{n\geq 0}} k^{-n} b^{(n)} =
\sum_{{n\geq 0}} \sum_{i,j=1}^N k^{-n} b^{(n)}_{ij} E_{ij}
\label{serie-b}
\ee
Plugging this expression into (\ref{rbrb}) leads to the following
commutation relations:
\bea
&&\sum_{m,n=0}^\infty [b^{(m)}_{1}, b^{(n)}_{2}]\, k_{1}^{-m}k_2^{-n}
\ =\   \sum_{p,q,r=0}^\infty\left\{\rule{0ex}{1.6em} \right.
   ig\left(\rule{0ex}{1.4em}b_{2}^{(q)}b_{1}^{(p)}P-
Pb_{1}^{(p)}b_{2}^{(q)}\right.+\\
&& \left.+(-1)^r\left(b_{2}^{(q)}Pb_{1}^{(p)}-
b_{1}^{(p)}Pb_{2}^{(q)}\right)\rule{0ex}{1.4em}\right)k_{1}^{-p-r-1}k_{2}^{r-q}
\left. -\, g^2\, \left[b_{2}^{(q)},b_{2}^{(p)}\right]
\, k_{1}^{-p-2r-2}k_{2}^{2r-q}
\rule{0ex}{1.6em}\right\}\nonumber
\eea
In order the $\salg$ algebra to be well-defined, no positive power of
$k$ can be admitted on the r.h.s. of the above expression. A direct
computation shows that spurious terms are avoided by the
necessary and sufficient conditions:
\be
b(k)b(-k) = f(k)\, \II\mb{with} f(k)=1+\sum_{m=1}^\infty
f_{2m}k^{-2m}\mb{even function}\label{consist}
\ee
\be
{[b_{1}^{(0)},b_{2}(k)]}=0 \mb{i.e. all the $b_{ij}^{(0)}$ generators
are central}\label{b0cent}
\ee
Remark that the condition (\ref{consist}) is a natural generalization of the
physical condition (\ref{rc}) obtained in section \ref{sect-Balg}.
In the following, we will take $f(k)$ as a pure \CC-function. The
normalization of $f(k)$ is fixed by $\lim_{k\rightarrow\infty}f(k)=1$.

Before commenting on these constraints, let us note that
the commutation relations can then be rewritten as:
\bea
[b^{(m)}_{1}, b^{(n)}_{2}] &=& ig\sum_{r=0}^{m-1}\left\{\rule{0ex}{1.6em}
   \left(b_{2}^{(n+r)}b_{1}^{(m-1-r)}-
b_{2}^{(m-1-r)}b_{1}^{(n+r)}\right)P\right.+\nonu
&&\left. +(-1)^r\left(b_{2}^{(n+r)}Pb_{1}^{(m-1-r)}-
b_{1}^{(m-1-r)}Pb_{2}^{(n+r)}\right)\rule{0ex}{1.6em} \right\}+\nonu
&&-g^2\, \sum_{r=0}^{\mu}\ [b_{2}^{(n+2r)},b_{2}^{(m-2-2r)}]
\mb{where} \mu=\left[\frac{m-2}{2}\right]
\eea
that is
\bea
[b^{(m)}_{ij}, b^{(n)}_{kl}] &=& ig\sum_{r=0}^{m-1}\left\{\rule{0ex}{1.6em}
   \left(b_{kj}^{(n+r)}b_{il}^{(m-1-r)}-
b_{kj}^{(m-1-r)}b_{il}^{(n+r)}\right)\right.+\nonu
&&\left. +(-1)^r\left(\delta_{il}\,b_{ka}^{(n+r)}b_{aj}^{(m-1-r)}-
\delta_{kj}\,b_{ia}^{(m-1-r)}b_{al}^{(n+r)}\right)\rule{0ex}{1.6em}
\right\}+\nonu
&&-g^2\, \sum_{r=0}^{\mu}\ \delta_{ij}\,\left(
b_{ka}^{(n+2r)}b_{al}^{(m-2-2r)}
-b_{ka}^{(m-2-2r)}b_{al}^{(n+2r)}\right)
\eea
and also:
\bea
[b_{ij}(k_1), b_{kl}(k_2)] &=& \frac{ig}{k_1-k_2}
\left(\rule{0ex}{1.3em}b_{kj}(k_2)b_{il}(k_1)
-b_{kj}(k_1)b_{il}(k_2)\right)+\nonu
&&+\frac{ig}{k_1+k_2}\left(\rule{0ex}{1.3em}
\delta_{il}\,b_{ka}(k_2)b_{aj}(k_1)-
\delta_{kj}\,b_{ia}(k_1)b_{al}(k_2)\right)+\nonu
&&-\frac{g^2}{k_1^2-k_2^2}\, \delta_{ij}\,\left(\rule{0ex}{1.3em}
b_{ka}(k_2)b_{al}(k_1)-b_{ka}(k_1)b_{al}(k_2)\right)
\eea
In particular, for $m=1$, we get
\be
[b^{(1)}_{ij}, b^{(n)}_{kl}] = ig\left(
b_{kj}^{(n)}b_{il}^{(0)}+\delta_{il}\,b_{ka}^{(n)}b_{aj}^{(0)}-
b_{kj}^{(0)}b_{il}^{(n)}-\delta_{kj}\,b_{ia}^{(0)}b_{al}^{(n)}\right)
\ee
which shows that in each representation where $b^{(0)}$ is a
constant, the $b^{(1)}$ generators form a Lie subalgebra and the
$b^{(n)}$ generators (for any given $n$) fall into representations
of this Lie subalgebra. In what follows, we will consider only this
type of representations.
\subsubsection{Commutative subalgebras of the $\salg$-algebra}
We give in this short section some commutative subalgebras which will
be used in the following.
\begin{prop}\label{prop.trb}
     Let us introduce
     \be
     t(k)=tr\Big( b(k)\Big)=\sum_{n=0}^\infty k^{-n}t^{(n)}
     \label{def.t}
     \ee
     $t(k)$ defines a commutative subalgebra of $\salg$: $[t(k_1),t(k_2)]=0$.
\end{prop}
\prf
We take the trace in the auxiliary space 1 of the relation
(\ref{com.bubv}):
\be
{[t(k_1),b(k_2)]}=\frac{2ik_1g}{k_1^2-k_2^2}\, [b(k_2),b(k_1)]
\label{com.tb}
\ee
Then, taking again the trace we get
\be
{[t(k_1),t(k_2)]}=\frac{2ik_1g}{k_1^2-k_2^2}\, tr\Big([b(k_2),b(k_1)]\Big)
\ee
 From ${[t(k_1),t(k_2)]}=-{[t(k_2),t(k_1)]}$, we deduce
$tr\Big([b(k_2),b(k_1)]\Big)=0$, which implies ${[t(k_1),t(k_2)]}=0$.
\finprf
\begin{prop}\label{prop.trMb}
     Let $\MM$ be any constant matrix such that $\MM^2=\II$
     and correspondingly
     \be
     \wt{t}_M(k)=tr\Big( \MM b(k)\Big)=\sum_{n=0}^\infty k^{-n}\wt{t}_M^{(n)}
     \label{def.wht}
     \ee
     $\wt{t}_M(k)$ defines a commutative subalgebra of $\salg$ which
     commutes with the one given in proposition \ref{prop.trb}:
     \be
     [\wt{t}_M(k_1),\wt{t}_M(k_2)]=0\mb{and} [t(k_1),\wt{t}_M(k_2)]=0
     \ee
\end{prop}
  \prf
We start again with (\ref{com.bubv}), multiply it by
$\MM_{1}\equiv\MM\otimes\II$, and then take the trace in the auxiliary space 1:
\bea
{[\wt{t}_M(k_1),b(k_2)]} &=& \frac{ig}{k_1-k_2}\, \Big(b(k_2)\MM b(k_1)
-b(k_1)\MM b(k_2)\Big)+\nonu
&&+
\frac{ig}{k_1+k_2}\, [b(k_2)b(k_1),\MM]+\nonu
&&-\frac{g^2\, tr(\MM)}{k_1^2-k_2^2}\, [b(k_2),b(k_1)]
\label{com.whtb}
\eea
Then, after a product on the left by $\MM$ and the trace, we get
\beano
{[\wt{t}_M(k_1),\wt{t}_M(k_2)]} &=& \frac{ig}{k_1-k_2}\,
tr\MM \Big(b(k_2)\MM b(k_1)-\MM b(k_1)\MM b(k_2)\Big)+\\
&& -\frac{g^2\, tr(\MM)}{k_1^2-k_2^2}\, tr\Big(\MM [b(k_2),b(k_1)]\Big)
\eeano
The left hand side of the above expression is skew-symmetric in $(k_1,k_2)$,
while the right hand side is symmetric, so that each sides are
identically zero. This proves the first part of the assertion.

Now, taking the trace of (\ref{com.whtb}), one gets
\be
{[\wt{t}_M(k_1),{t}(k_2)]}=\frac{ig}{k_1-k_2}\, tr\Big(b(k_2)\MM 
b(k_1)-b(k_1)\MM b(k_2)\Big)
\ee
which proves that ${[\wt{t}_M(k_1),{t}(k_2)]}$ is symmetric under the
exchange $u\leftrightarrow v$. On the other hand, if one multiplies
(\ref{com.tb}) by $\MM$ and then take the trace, one obtains
\be
{[{t}(k_1),\wt{t}_M(k_2)]}=\frac{2ik_1g}{k_1^2-k_2^2}\, tr\Big(\MM 
[b(k_2),b(k_1)]\Big)
\ee
The antisymmetric part in $(k_1,k_2)$ of the above expression shows that
we have $tr(\MM [b(k_2),b(k_1)])=0$, which then proves the second assertion.
\finprf

\subsection{Analysis of the consistency relations}
We now turn to the consequences of the constraints
(\ref{consist}-\ref{b0cent}). As far as $b^{(0)}$ is concerned, we
have already seen that it must be central, but the consistency
relation (\ref{consist}) also implies
\be
\left(b^{(0)}\right)^2=\II\label{b02=1}
\ee
In the Fock representations $\brep$ we consider, $b^{(0)}$
is a constant matrix
and (\ref{b02=1}) shows that it can be diagonalized (by a constant
$gl(N)$ matrix). Thus, up to a conjugation, one can suppose that we
have
\be
b^{(0)}=\EE=\sum_{i=1}^N\, \eps_{i}\, E_{ii}\mb{with} \eps_{i}=\pm1
\label{b0=E}
\ee
For convenience, we will use the notation
\be\begin{array}{l}
A=\{\alpha\mb{such that}\eps_{\alpha}=+1\}\subset [1,N]
\mb{;}\mbox{dim}A=M\\
\bar{A}=\{\baral\mb{such that}\eps_{\baral}=-1\}\subset [1,N]
\mb{;}\mbox{dim}\bar{A}=N-M
\end{array}
\ee
In the basis (\ref{b0=E}), the commutation relations with the Lie subalgebra
generators read
\be
[b^{(1)}_{ij}, b^{(n)}_{kl}] = ig(\eps_{i}+\eps_{j})\left(\delta_{il}\,
b_{kj}^{(n)}- \delta_{kj}\,b_{il}^{(n)}\right)
\ee
which indicates that the Lie subalgebra depends on the choice of $\EE$.
Indeed, the
consistency relation (\ref{consist}) together with the choice of
$b^{(0)}=\EE$ lead to
\be
b^{(1)}_{\alpha\barbt}=0=b^{(1)}_{\baral \beta}
\ee
so that the Lie subalgebra in $\salg$ is a $gl(M)\oplus gl(N-M)$ one.

More generally, the analysis of the consistency
relations shows that $b^{(2n+1)}_{\alpha\barbt}$,
$b^{(2n+1)}_{\barbt\alpha}$,
$b^{(2n+1)}_{\alpha\beta}$ and $b^{(2n+1)}_{\baral\barbt}$
can be expressed in
terms of the other generators, so that
$\salg$ is generated by:
\bea
\salg^{(2n)} =\left\{b^{(2n)}_{\alpha\barbt},\
b^{(2n)}_{\barbt\alpha}\right\}_{(\alpha\in A,\,\barbt\in\bar{A})}\ =\
(M,\overline{N-M})+(\overline{M},N-M) &&\\
\salg^{(2n+1)} =\left\{b^{(2n+1)}_{\alpha\beta},
\ b^{(2n+1)}_{\baral\barbt}
\right\}_{(\alpha,\beta\in A,\,\baral,\barbt\in\bar{A})}\ =\
(M^2,0)+(0,(N-M)^2) &&
\eea
where we have indicated the decomposition in $gl(M)\oplus gl(N-M)$
representations.

\sect{The NLS reflection matrices \label{sect-boundcond}}
Now, we come to the explicit construction of the reflection matrices
as defined by (\ref{B}-\ref{bHa}). As already expressed in section 
\ref{sect-Balg}, to
each allowed $B$ matrix (defined up to a $gl(N)$ conjugation, see
below) is associated a Fock space representation $\cf_B$ of the reflection
algebra $(\balg, I)$.
Indeed the value of the operator $b(k)$ on the Fock space vacuum $\Omega$
is directly given by the matrix $B(k)$: see (\ref{bO}).

\subsection{Classification of the $B$-reflection matrices}
Expliciting the condition (\ref{RBRB}) with $R$ again defined by
(\ref{truc}) leads to the equations:
\be
  {[B(k_1),B(k_2)]}=0 \label{comBB}
  \ee
  \be
  B_1(k_1)B_1(k_2)-B_2(k_2)B_2(k_1)=
\frac{k_1+k_2}{k_1-k_2}\left( \rule{0ex}{1.3em}
B_2(k_2)B_1(k_1)-B_2(k_1)B_1(k_2)\right) \label{eqB}
\ee
 From (\ref{comBB}), one immediately deduces that the $B(k)$ matrices
($\forall\ k$) can simultaneously put into a triangular form using
a constant $Gl(N)$ matrix:
\be
B(k)=U\, T(k)\, U^{-1}\mb{with} [T(k_{1}),T(k_2)]=0\label{triang}
\ee
Moreover, since we have
\be
R_{12}(k_1-k_2)U_1U_2=U_2U_1R_{12}(k_1-k_2)
\ee
the transformation (\ref{triang}) defines an automorphism of the
whole algebra $\balg$, as well as of the condition (\ref{rc}).
Hence, we can suppose without any restriction that the matrices
$B(k)$ are triangular.

Imposing $k_2=-k_1$ in (\ref{eqB}) implies
\be
B(k)B(-k)=B(-k)B(k)=\rho(k)\II
\ee
where $\rho$ is an even function which must be real in order to
satisfy (\ref{bHa}).

Then, a detailed study of (\ref{eqB}) lead to the following classification
of $B$ reflections matrices:
\bea
\mbox{{\bf Case $\rho(k)\neq0$}}
&& B(k)=\beta(k)\frac{\II+iak\, \EE}{1+iak}\label{clas1}\\
&& B(k)=\beta(k)\,\EE \\
&& B(k)=\beta(k)\left(\II+ak\JJ\right)\label{clas3}\\[2.3ex]
\mbox{{\bf Case $\rho(k)=0$}}
&& B(k)={\beta(k)}\,\JJ\label{clas4}
\eea
with the conditions
\be
\EE^2=\II\mb{;}\JJ^2=0\mb{;}a\in\RR
\ee
Note that because of the $Gl(N)$ invariance,  $\EE$ can be taken diagonal:
\be
\EE=\sum_{i=1}^N \eps_i\, E_{ii}\mb{with}
(E_{ii})_{kl}=\delta_{ik}\delta_{il}
\ee
and $\JJ$ can be fixed to its Jordanian form:
\be
\JJ=\left(\begin{array}{cccc}
J_1&0&\cdots&0\\
0&J_2&\ddots&\vdots\\
\vdots&\ddots&\ddots&0\\
0&\cdots&0&J_r
\end{array}\right)
\mb{with} J_i=(0)\mb{or}\left(\begin{array}{cc} 0&1\\
0&0\end{array}\right)\ 1\leq i\leq r
\ee
Requiring the matrix $B(k)$ to obey the condition (\ref{bHa}) leads to
the supplementary relation:
\be
\beta(k)^*=\beta(-k)
\ee
Let us remark that the solution (\ref{clas3}) obey to (\ref{RBRB}), but not to
the development (\ref{eq:bseries}), and as such should be rejected in 
this context.
The other solutions can all be expanded in series of $k^{-1}$
(provided $\beta(k)$ can be).

\subsection{Boundary conditions associated to $B(k)$}
 From the algebraic study of the $\salg$ algebra and its Fock
representations, one can determine the boundary conditions
obeyed by the physical fields $\Psi(x,t)$. For such a purpose
we associate to each reflection matrix $B(k)$ a differential
operator (in the variable $x$) $D_B$ that will act on $\Psi(x,t)$.
 From (\ref{gbc}), one infers that this operator will be
fixed by its value on the eigenstates
\be
\lambda_k(x)=e^{ikx}+B(k)e^{-ikx}
\ee
through the equation
\be
\left. D_B\, \lambda_k(x)\right|_{x=0}=0 \label{DB}
\ee
We restrict our analysis to the physical condition
\be
B(k)\cdot B(-k)=\II
\ee
In all cases, the boundary condition for the physical field $\Psi(x,t)$
takes the form:
\be
\left. D_B\, \Psi(x,t)\right|_{x=0}=0
\ee
We specify below the operator $D_{B}$ according to the
classification (\ref{clas1}-\ref{clas4}).

\subsubsection{Case of $B(k)=\beta(k)\II$}
 From the condition $\beta(k)\beta(-k)=1$, one deduces that $\beta(k)$
can be rewritten as
\be
\beta(k)=\frac{A(k)+i}{A(k)-i} \mb{with} A(k)\mbox{ real odd function}
\label{defA}
\ee
Defining the differential operator $D_A$ through
\be
(D_Af)(y)=\int dx\ \hat{A}(x)\, f(x+y)\ \forall\, f\mb{\CC-function}
\label{defDA}
\ee
where $\hat{A}$ is the inverse Fourier transform of $A$. By definition
of $\hat{A}$, we have
\be
D_A e_p=A(p) e_p\mb{where} e_p(y)=\exp(ipy)
\ee
This implies
\be
(D_A+i)\lambda_k=(A(k)+i)e_k+\beta(k)(A(-k)+i)e_{-k}
=(A(k)+i)(e_k-e_{-k})
\ee
In other words, one gets $(D_A+i)\lambda_k(x)|_{x=0}=0$, and
$D_B=D_A+i$ is the required operator.

Let us remark that in the particular case evocated in (\ref{SchB})
\be
\beta(k)=\frac{k-i\eta}{k+i\eta}\ \ i.e.\ \ A(k)=\frac{k}{\eta}
\ee
we get $\hat{A}(x)=\frac{-i}{\eta}\frac{\prt}{\prt x}$
and we recover the boundary condition (\ref{bc0}), using the fact that
$\prt_xf(x+y)=\prt_yf(x+y)$.

\subsubsection{Case of $B(k)=\beta(k)\EE$}
As in the previous case, we introduce the odd function $A$
defined in (\ref{defA}). The calculation is similar to the previous one,
except for the matrix dependence. We define the differential
operator $D_A$ through (\ref{defDA}).  Then, one can choose
\be
D_B=\frac{\II+\EE}{2}(D_A+i)+\frac{\II-\EE}{2}(D_A+i)\prt_x
\ee
This operator will obey the condition (\ref{DB}), because of the
properties:
\bea
D_Ae_k &=& A(k)e_k \mb{and} D_A\prt_xe_k\ =\ ikA(k)e_k\\
(\II+\EE)\EE &=& \II+\EE \mb{and} (\II-\EE)\EE \ =\ -(\II-\EE)
\eea
which lead to
\be
D_B\lambda_k = \frac{\II+\EE}{2}(A(k)+i)(e_k-e_{-k})
+ \frac{\II-\EE}{2}ik(A(k)+i)(e_k-e_{-k})
\ee
It is then obvious that $D_B\lambda_k|_{x=0} =0$.

\subsubsection{Case of $B(k)=
\beta(k)\frac{\displaystyle\II+iak\,\EE}{\displaystyle1+iak}$}
This case comes as a mixing of the two previous ones.
We first rewrite $B$ as:
\be
B(k)=\beta(k)\frac{\II+\EE}{2}+\wt{\beta}(k)\frac{\II-\EE}{2}
\mb{with} \wt{\beta}(k)=\beta(k)\frac{1-iak}{1+iak}
\ee
and define $A(k)$ by (\ref{defA}) and $\wt{A}(k)$ by
\be
\wt{\beta}(k)=\frac{\wt{A}(k)+i}{\wt{A}(k)-i}
\mb{i.e.} \wt{A}(k)=\frac{{A}(k)+ak}{1-akA(k)}
\ee
The differential operators $D_A$ and $D_{\wt A}$ will be
constructed as above, and it is straightforward to check that
\be
D_B=\frac{\II+\EE}{2}(D_A+i)+\frac{\II-\EE}{2}(D_{\wt A}+i)
\ee
obeys the relation (\ref{DB}).
\section{Spontaneous symmetry breaking\label{sect-symbreak}}
In the Fock space representation $\cf_B$ of the reflection
algebra $\{\balg, I\}$, once the matrix $B(k)$ is fixed, one knows all
the operators $b^{(n)}$ which have non-vanishing value on $\Omega$.
Since the $\salg$-algebra constitutes the symmetry algebra of our
problem, we are exactly faced with a mechanism of spontaneous
symmetry breaking for our reflection algebra, itself part of our
$\balg$-algebra.
We present a generating function
for the broken generators, and show that these generators form a
commutative subalgebra of $\salg$.
\subsection{Case of $B(k)=\beta(k)\II$}
We start with the reflection matrix:
\be
B(k)=\beta(k)\II \mb{with} \beta(k)=\sum_{n\in J}\beta_{n}k^{-n}
\mb{;} \beta_{n}\neq0\ \forall n\in J\subset\NN
\ee
  The vacuum expectation value of $b(k)$ is thus
  \be
  <b_{ij}^{(p)}> = \beta_{p} \delta_{ij}\
  \Rightarrow\ <b_{ii}^{(p)}-b_{jj}^{(p)}>=0\
  \forall p
  \ee
Thus, the broken generators are all contained in\footnote{We will
ignore in our general approach the possibility for two broken
generators of different grade, i.e. $trb^{(n)}$ and $trb^{(m)}$, to
combine and form a third element, i.e.
$\beta_{m}trb^{(n)}-\beta_{n}trb^{(m)}$, for which the symmetry is
restaured. A way of determining whether this last element (which
algebraically annihilates the vacuum) would be to construct, when
possible, a compensating transformation in the momentum space which
preserves the correlation functions.}:
\be
   \sum_{i=1}^{N} b_{ii}^{(p)}=tr(b^{(p)})\ \forall p\in J \label{brok1}
\ee
We first gather them into
\be
t_{B}(u)=\sum_{p\in J}u^{-p} tr(b^{(p)})
\ee
   Let us define the operator (depending on the new variable $u$ and on
   the variable $x=k^{-1}$):
   \be
   D^k_u=\sum_{p\in J}\frac{1}{p!} \left(u^{-1}\prt\right)^{p}
   \equiv d(u^{-1}\prt)
   \mb{with} \prt=\frac{\prt}{\prt(k^{-1})}
   \mb{;} d(x)=\sum_{p\in J} \frac{x^p}{p!}\label{defDku}
   \ee
   Then, from an obvious calculation, we have:
   \begin{prop}
   The generating function $t_{B}(u)$ is given by
   \be
   t_{B}(u)=\left.\mbox{tr}(D^k_ub(k))\right|_{k=\infty}=
   \left. D^k_u\,t(k)\right|_{k=\infty}
   \ee
   where $t(k)$ has been defined in (\ref{def.t}). It satisfies
   \be
   t_{B}(u)\Omega=B(u)\Omega
   \ee
   \end{prop}
  \subsection{Case of $B(k)=\beta(k)\EE$}
  Similarly, for
   \be
   B(k)=\beta(k)\EE\mb{with} \left\{\begin{array}{l}
   \displaystyle \beta(k)=\sum_{n\in J}\beta_{n}k^{-n}
   \mb{;} \beta_{n}\neq0\ \forall n\in J\subset\NN\\[2.3ex]
   \displaystyle
   \EE=\sum_{\alpha=1}^ME_{\alpha\alpha}-\sum_{\baral=M+1}^N
   E_{\baral\baral}=\sum_{i=1}^N \eps_{i}E_{ii}
   \end{array}\right.
   \ee
   we get the conditions:
  \be
  <b_{ij}^{(p)}> = \beta_{p}\eps_{i} \delta_{ij}\ \Rightarrow\
  <\eps_{i}b_{ii}^{(p)}-\eps_{j}b_{jj}^{(p)}>=0\ \forall p
  \ee
Then, the broken generators are of the form:
  \be
   \sum_{i=1}^{N} \eps_{i}b_{ii}^{(p)}=tr(\EE b^{(p)})\ \forall p\in J
   \label{brok2}
\ee
With the same calculation, as in previous section, we obtain:
\begin{prop}
The generating function for the broken generators is given by:
   \be
  t_{B}(u)=\sum_{p\in J}u^{-p} tr(\EE b^{(p)})=
  \mbox{tr}\left.(\EE D^k_ub(k))\right|_{k=\infty}=
    \left. D^k_u\,\wt{t}_{E}(k)\right|_{k=\infty}\ \forall u
   \ee
  with the same definition (\ref{defDku}) of $D^k_u=d(u^{-1}\prt)$.
    $\wt{t}_{E}(k)$ has been defined in (\ref{def.wht}), with here $\MM=\EE$.
    $t_{B}(u)$ satisfies
   \be
   t_{B}(u)\Omega=B(u)\Omega
   \ee
  \end{prop}

\subsection{General case}
Up to a redefinition of $\beta$ and $a$, one can rewrite $B(k)$
defined in (\ref{clas1}) as:
   \be
B(k)=\frac{\beta(k)}{1+a}(\EE+ak^{-1}\II)
\ee
so that the two previous cases are given by the limits
$a\rightarrow0$ or $\infty$.
We have
\be
  <b_{ij}(k)>=\frac{\beta(k)}{1+a}(\eps_{i}+ak^{-1})\delta_{ij}
\ee
which shows that
\bea
&&<b_{\alpha\alpha}(k)-b_{\beta\beta}(k)>=0,\ \ \forall\, 
\alpha,\beta=1,\dots,M
\label{bii}\\
&&<b_{\baral\baral}(k)-b_{\barbt\barbt}(k)>=0,\ \ \forall\,
\baral,\barbt=M+1,\dots,N \label{bbibi}
\eea
Once again, the broken generators are gathered in a generating
function $t_{B}(u)$, and we get
\begin{prop}
The generating function for broken generators is given by
\be
t_{B}(u)=\mbox{tr}\left.\left(D^k_ub(k)\right)\right|_{k=\infty}=
\frac{1}{1+a}\left.\left( d(u^{-1}\prt)t(k)+a\wh{d}(u^{-1}\prt)\wt{t}_{E}(k)
\right)\right|_{k=\infty}
\ee
where now the matrix differential operator reads:
\[
D^k_{u} = \frac{1}{1+a}\left( \EE\,d(u^{-1}\prt)+
\II\, a\wh{d}(u^{-1}\prt)\right)
\mb{with}\beta(k)=\sum_{p\in J} \beta_p k^{-p}
\]
As above
\be
d(x)=\sum_{p\in J} \frac{1}{p!} x^p \mb{and}
\wh{d}(x)=\sum_{p\in J} \frac{1}{(p+1)!} x^{p+1}
\ee
    $t_{B}(u)$ satisfies
   \be
   t_{B}(u)\Omega=B(u)\Omega
   \ee
\end{prop}
\begin{coro}
The "broken generators" form a commutative algebra:
\be
[t_{B}(u),t_{B}(v)]=0
\ee
\end{coro}
\prf
Direct consequence of properties (\ref{prop.trb}) and (\ref{prop.trMb}).
\finprf

{\bf Remark:} In all cases, $t_{B}(u)$ is a scalar matrix, which
means that we have at most only one broken generator at each level.

\subsection{Example: $B(k)=\II$}
We are in the particular case $\eta=0$
of the conditions given in (\ref{bc0}-\ref{bcinf}), so that the 
boundary conditions
are here
\be
\lim_{x\rightarrow\infty}\Phi(x,t)=0 \mb{and}
\lim_{x\downarrow0}\prt_{x}\Phi(x,t)=0
\ee
Due to the form of $B(k)$, it is obvious that none of the generators
is broken. Hence, we are in the very specific situation where no spontaneous
symmetry breaking occurs.

However, remark that the boundary has an effect on the symmetry
algebra: indeed, the Yangian symmetry which appears in the NLS
hierarchy on the full line \cite{MRSZ}
is reduced here down to the `smaller' $\salg$-algebra.

\section{The particular case $N=2$ \label{n=2}}
\subsection{Characteristic of $N=2$}
The distinctive feature of $N=2$ lies in the fact that the Pauli
matrices, together with the identity matrix, form a basis of
$2\times2$ matrices and that all these matrices obey to the property
\ref{prop.trMb}. Indeed, one has:
\begin{prop}
     For $N=2$, $t(k)$ is central in the $\salg$-algebra, and
     a generating system of this algebra is given by
\be
t(k)=tr\, b(k)=b_{11}(k)+b_{22}(k) \mb{and}
\wt{t}_{a}(k)=tr\big(\sigma_{a}b(k)\big),\ a=1,2,3
\ee
or more explicitly
\[
\wt{t}_{1}(k)=b_{12}(k)+b_{21}(k)\ ;\
\wt{t}_{2}(k)=i\Big(b_{12}(k)-b_{21}(k)\Big)
\ ;\ \wt{t}_{3}(k)=b_{11}(k)-b_{22}(k)
\]
\end{prop}
\prf
$\wt{t}_{a}(k)$ are all generators obeying the property \ref{prop.trMb}, which
proves that $t(k)$ commutes with these generators. It is then enough
to show that these generators form a generating system (together with 
$t(k)$) to
proves that $t(k)$ is central (using also property \ref{prop.trb}).
Using
\be
\sigma_{a}\sigma_{b}=\delta_{ab}\,\II+i{\eps_{ab}}^c\,\sigma_{c}
\label{sigsig}
\ee
and $tr\sigma_{a}=0$, one gets
\be
b(k)=\half\Big(t(k)+\sum_{a=1}^3\, \wt{t}_{a}(k)\, \sigma_{a}\Big)
\label{eq:btta}
\ee
showing therefore that $\{t(k),\wt{t}_{a}(k),\ a=1,2,3\}$ is a
generating system (as $b_{ij}(k)$).
\finprf
The above property allows us to give simpler commutation relation for
the $\salg$-algebra:
\begin{prop}
     The commutation relations of the $\salg$ algebra, for $N=2$, are
     given by
     \bea
     {[t(k_1),t(k_2)]} &=&0 \ ;\ [t(k_1),\wt{t}_{a}(k_2)]\ =\ 0
     \label{eq:t.cent}\\
     {[\wt{t}_{a}(k_1),\wt{t}_{b}(k_2)]} &=& \frac{-2g{\eps_{ab}}^c}{k_1-k_2}\,
  \frac{k_1\,{t}(k_1)\wt{t}_{c}(k_2)-k_2\,{t}(k_2)\wt{t}_{c}(k_1)}{k_1+k_2+ig}
  \label{eq:tatb}\\
  \wt{t}_{a}(k_1)\wt{t}_{b}(k_2) &=&\wt{t}_{a}(k_2)\wt{t}_{b}(k_1)
  \label{eq:tt=tt}
     \eea
  \end{prop}
  \prf
  The equations (\ref{eq:t.cent}) are just the rephrasing of previous
  property. It remains to proves (\ref{eq:tatb}). We start with
  (\ref{com.whtb}), for $\MM=\sigma_{a}$, multiply it by $\sigma_{b}$ and take
  the trace:
  \bea
  {[\wt{t}_{a}(k_1),\wt{t}_{b}(k_2)]} &=& \frac{ig}{k_1-k_2}\ tr\Big(\,
  \sigma_{b}b(k_2)\sigma_{a}b(k_1)-\sigma_{b}b(k_1)\sigma_{a}b(k_2)\,\Big)+\nonu
  &&+\frac{ig}{k_1+k_2}\ tr\Big(\,
  \sigma_{b}\big[b(k_2)b(k_1)\,,\,\sigma_{a}\big]\,\Big)
  \eea
  Using the expression (\ref{eq:btta}), this can be rewritten as
  \bea
  4{[\wt{t}_{a}(k_1),\wt{t}_{b}(k_2)]} &=&\frac{ig}{k_1-k_2}\
  tr\big(\, \sigma_{b}\sigma_{d}\sigma_{a}\sigma_{c}\,\big)
  \Big(\,\wt{t}_{d}(k_2)\wt{t}_{c}(k_1)-
  \wt{t}_{d}(k_1)\wt{t}_{c}(k_2)\,\Big)+\nonu
&& +\frac{ig}{k_1+k_2}tr\big(\, 
\sigma_{b}[\sigma_{c}\sigma_{d},\sigma_{a}]\,\big)
  \,\wt{t}_{c}(k_2)\wt{t}_{d}(k_1)+\nonu
  &&+4i{\eps_{ab}}^c\,
  \Big(\,\frac{ig}{k_1+k_2}\,\big({t}(k_1)\wt{t}_{c}(k_2)
  +{t}(k_2)\wt{t}_{c}(k_1)\big)+\nonu
  &&\hspace{3.2em}+\frac{ig}{k_1-k_2}\,\big({t}(k_1)\wt{t}_{c}(k_2)-
  {t}(k_2)\wt{t}_{c}(k_1)\big)\,\Big)\label{tatb-1}
  \eea
  with summation over repeated indices.
A direct calculation shows
\bea
&&tr\Big(\,
\sigma_{b}[\sigma_{c}\sigma_{d},\sigma_{a}]\,\Big)=
4\Big(\delta_{bc}\delta_{ad}-\delta_{ac}\delta_{bd}\Big)\\
&&tr\Big(\,
\sigma_{b}\sigma_{d}\sigma_{a}\sigma_{c}\,\Big)=
2\Big(\delta_{bd}\delta_{ac}+\delta_{ad}\delta_{bc}-\delta_{ab}\delta_{cd}\Big)
\eea
so that we get
\beano
  {[\wt{t}_{a}(k_1),\wt{t}_{b}(k_2)]} &=&-\half\frac{ig}{k_1-k_2}\
  \Big(\, 
[\wt{t}_{a}(k_1),\wt{t}_{b}(k_2)]-
[\wt{t}_{a}(k_2),\wt{t}_{b}(k_1)]\,\Big)+\\
&& +\frac{ig}{k_1+k_2} \Big(\, 
\wt{t}_{b}(k_2)\wt{t}_{a}(k_1)-\wt{t}_{a}(k_2)\wt{t}_{b}(k_1)\,\Big)+\\
  &&+\frac{i{\eps_{ab}}^c}{k_1^2-k_2^2}\,
  \Big(\,2igk_1\,{t}(k_1)\wt{t}_{c}(k_2)-
2igk_2\,{t}(k_2)\wt{t}_{c}(k_1)\,\Big)
  \eeano
Exchanging $(a,k_1)\leftrightarrow (b,k_2)$, the above equality leads to
\beano
  {[\wt{t}_{b}(k_2),\wt{t}_{a}(k_1)]} &=&-\half\frac{ig}{k_1-k_2}\
  \Big(\, 
[\wt{t}_{a}(k_1),\wt{t}_{b}(k_2)]
-[\wt{t}_{a}(k_2),\wt{t}_{b}(k_1)]\,\Big)+\\
&& +\frac{ig}{k_1+k_2} \Big(\,
\wt{t}_{a}(k_1)\wt{t}_{b}(k_2)-\wt{t}_{b}(k_1)\wt{t}_{a}(k_2)\,\Big)+\\
  &&-\frac{i{\eps_{ab}}^c}{k_1^2-k_2^2}\,
  \Big(\,2igk_1\,{t}(k_1)\wt{t}_{c}(k_2)-
2igk_2\,{t}(k_2)\wt{t}_{c}(k_1)\,\Big)
  \eeano
  so that
\beano
  2{[\wt{t}_{a}(k_1),\wt{t}_{b}(k_2)]} &=& {[\wt{t}_{a}(k_1),\wt{t}_{b}(k_2)]}-
  {[\wt{t}_{b}(k_2),\wt{t}_{a}(k_1)]} \\
  &=&\frac{ig}{k_1+k_2}\
  \Big(\, 
[\wt{t}_{b}(k_2),\wt{t}_{a}(k_1)]+
[\wt{t}_{b}(k_1),\wt{t}_{a}(k_2)]\,\Big)+\\
  &&-\frac{i{\eps_{ab}}^c}{k_1^2-k_2^2}\,
  \Big(\,4igk_1\,{t}(k_1)\wt{t}_{c}(k_2)-
4igk_2\,{t}(k_2)\wt{t}_{c}(k_1)\,\Big)
  \eeano
  Computing
  $[\wt{t}_{b}(k_2),\wt{t}_{a}(k_1)]+[\wt{t}_{b}(k_1),\wt{t}_{a}(k_2)]$ 
leads to
  \beano
  &&\frac{k_1-k_2+ig}{k_1-k_2}\Big([\wt{t}_{a}(k_1),\wt{t}_{b}(k_2)]+
  [\wt{t}_{a}(k_2),\wt{t}_{b}(k_1)]\Big) =\\
  &&=\frac{4i{\eps_{ab}}^c}{k_1^2-k_2^2}\,
  \Big(\,igk_1\,{t}(k_1)\wt{t}_{c}(k_2)-igk_2\,{t}(k_2)\wt{t}_{c}(k_1)\,\Big)
  \eeano
  which proves (\ref{eq:tatb}).
 
On the other hand, starting from (\ref{tatb-1}), one computes
\[
  {[\wt{t}_{a}(k_1),\wt{t}_{b}(k_2)]} - {[\wt{t}_{a}(k_2),\wt{t}_{b}(k_1)]}  =
  \frac{ig}{k_1+k_2}\ \Big(\,
  \{\wt{t}_{a}(k_1),\wt{t}_{b}(k_2)\} -  \{\wt{t}_{a}(k_2),\wt{t}_{b}(k_1)\}
\Big)
\]
Then, using (\ref{eq:tatb}), one gets
\bea
{[\wt{t}_{a}(k_1),\wt{t}_{b}(k_2)]} &=& {[\wt{t}_{a}(k_2),\wt{t}_{b}(k_1)]}\\
  \{\wt{t}_{a}(k_1),\wt{t}_{b}(k_2)\} &=& \{\wt{t}_{a}(k_2),\wt{t}_{b}(k_1)\}
\eea
which is equivalent to the second relation.
\finprf
\begin{coro}
     The commutation relations with the Lie subalgebra generators read
     \bea
{[\wt{t}_{a}^{(1)},\wt{t}_{b}(k_2)]} &=& -2g{\eps_{ab}}^c\,
t^{(0)}\wt{t}_{c}(k_2)\\
{[{t}^{(1)},\wt{t}_{b}(k_2)]} &=& 0
    \eea
\end{coro}
\prf
One picks up the term $k_1^{-1}$ in the previous commutators.
\finprf

\subsection{Vacuum preserving algebra when $B(k)=\beta(k)\II$}
In this case, the broken generators are all gathered in $t(k)$, and
we can replace $t(k)$ (which is central) by its value $2\beta(k)$.

Thus, the vacuum preserving algebra is just the one generated by
$\wt{t}_{a}(k)$, $a=1,2,3$. Its commutation relations
read
\be
  {[\wt{t}_{a}(k_1),\wt{t}_{b}(k_2)]} 
=\frac{-4g{\eps_{ab}}^c}{(k_1+k_2+ig)(k_1-k_2)}\,
  \Big(\,k_1\,{\beta}(k_1)\wt{t}_{c}(k_2)-k_2\,{\beta}(k_2)
\wt{t}_{c}(k_1)\,\Big)
  \ee
Specifying to $m=n=1$, we get
\be
{[\wt{t}_{a}^{(1)},\wt{t}_{b}^{(1)}]} = -4g{\eps_{ab}}^c\,\wt{t}_{c}^{(1)}
\ee
which is an $sl_{2}$ Lie subalgebra.

Then, at the Lie subalgebra level, the spontaneous
symmetry breaking is a $gl_{2}\rightarrow sl_{2}$ one.

\subsection{Vacuum preserving algebra when $B(k)=\beta(k)\sigma_{3}$}
The broken generator is now $\wt{t}_{3}(k)$, and its value is
$2\beta(k)$.

Thus, the vacuum preserving
algebra is now the one generated by $t(k)$, $\wt{t}_{1}(k)$ and
$\wt{t}_{2}(k)$. It is indeed an algebra, as can be seen from the
commutators:
\beano
{[{t}{(k_1)},t{(k_2)}]} &=& 0 \qquad
{[t{(k_1)},\wt{t}_{a}{(k_2)}]} =0,\ a=1,2\\
{[\wt{t}_{1}{(k_1)},\wt{t}_{1}{(k_2)}]} &=& 0\qquad
{[\wt{t}_{2}{(k_1)},\wt{t}_{2}{(k_2)}]} =0\\
{[\wt{t}_{1}{(k_1)},\wt{t}_{2}{(k_2)}]} &=& \frac{-4g}{(k_1+k_2+ig)(k_1-k_2)}\,
  \Big(\,k_1\,{\beta}(k_2){t}(k_1)-k_2\,{\beta}(k_1){t}(k_2)\,\Big)
\eeano
However, since we are in the case $\EE=\sigma_{3}\neq\II$, the
condition $b(k)b(-k)=\II$ implies in particular
$b_{12}^{(1)}=b_{12}^{(1)}=0$, or in our generating system
$\wt{t}_{1}^{(1)}=\wt{t}_{2}^{(1)}=0$. Thus, for
the  values $m=n=1$, we get only one generator ${t}^{(1)}$,
and one recognizes a $gl_{1}$ Lie subalgebra.

Then, at the Lie subalgebra level, the spontaneous
symmetry breaking is a $gl_{1}\oplus gl_{1}\rightarrow gl_{1}$ one.

\sect{Outlook and Conclusions\label{sect-concl}}

We proposed in this paper an algebraic approach for studying
the symmetry content and the phenomenon of spontaneous symmetry
breaking in integrable systems on the half line. The main tool is
the quantum inverse scattering transform, where the familiar
Zamolodchikov-Faddeev (ZF) algebra is replaced by the the
boundary algebra $\balg $. The latter reflects the breakdown of translation
invariance on the half line, which is codified by a set of boundary
generators. These generators close a Sklyanin type subalgebra
${\cal S}_R\subset \balg $ and commute with the Hamiltonians of the whole
integrable hierarchy under consideration. For this reason ${\cal S}_R$
represents the central point of our investigation.

The basic and actually unique input of the scheme is the $R$-matrix,
which describes the two-body scattering and determines, via
the boundary Yang-Baxter equation, all reflection matrices
respecting integrability. Each reflection matrix in turn fixes a
boundary condition and therefore the dynamics and the symmetry
of the system.
The concrete example, we have focused on, is the $gl(N)$-NLS
model, whose $R$-matrix is given by (\ref{truc}). If one considers
this model on the whole line, the underlying symmetry algebra is
the Yangian $Y(sl(N))$
\cite{MW}. The corresponding generators can be constructed
\cite{MRSZ} in terms of the associated ZF algebra.
The latter admits unique (up to unitary equivalence) Fock
representation, whose vacuum is annihilated by all Yangian generators,
i.e. the whole Yangian $Y(sl(N))$ is an exact symmetry of the theory.
On the half line the situation is quite
different. The counterpart of $Y(sl(N))$ is the
Sklyanin algebra ${\cal S}_R$. The integrability
preserving boundary conditions are encoded in the reflection
matrices, which at the same time parametrize the inequivalent
Fock representations of the underlying boundary algebra $\balg $.
The existence of such representations is crucial.
It has no analog in the ZF algebra and
allows to describe the physically inequivalent phases of the
system on the half line. The generators of ${\cal S}_R$, which do not
annihilate the vacuum of a given phase, are spontaneously broken in
that phase.

The framework developed in this paper can be applied in a more general
context to the trigonometric and the elliptic series of $R$-matrices as well.
It will be interesting to investigate the structure of the associated
${\cal S}_R$ algebras, which will shed new light on the symmetry content
of the corresponding integrable models on the half line.


\begin{thebibliography}{99}

\bibitem{GS} S. Ghoshal and A. B. Zamolodchikov, Int. J. Mod. Phys.
{\bf A9} (1994) 3841.

\bibitem{FK} A. Fring and R. K\"oberle, Nucl. Phys. {\bf B419} (1994) 647;
{\it ibid.} {\bf B421} (1994) 159.

\bibitem{FS} P. Fendley and H. Saleur, Nucl. Phys.  {\bf B428} (1994) 681.

\bibitem{SSW} H. Saleur, S. Skorik and N. P. Warner,
Nucl. Phys. {\bf B441} (1995) 421.

\bibitem{BCDR} P. Bowcock, E. Corrigan, P. E. Dorey and R. H. Rietdijk,
Nucl. Phys.  {\bf B445} (1995) 469.

\bibitem{C1} E. Corrigan, Int. J. Mod. Phys.  {\bf A13} (1998) 2709.

\bibitem{LM} A. Liguori and M. Mintchev, Nucl. Phys.  {\bf B522} (1998) 345.

\bibitem{C2} E. Corrigan, {\sl Boundary bound states in integrable
quantum field theory}, proceedings of {\it Non-Perturbative Quantum
Effects 2000}, {\tt hep-th/0010094}.

\bibitem{AOS} I. Affleck, M. Oshikawa, H. Saleur, {\sl Boundary
critical phenomena in $SU(3)$ spin chains},
{\tt cond-mat/0011454}.

\bibitem {LMZ} A. Liguori, M. Mintchev and L. Zhao, Commun. Math. Phys.
{\bf 194} (1998) 569.

\bibitem{C} V. I. Cherednik, Theor. Math. Phys. {\bf 61} (1984) 977.

\bibitem{GLM1} M. Gattobigio, A. Liguori and M. Mintchev, Phys. Lett.
  {\bf B428}
(1998) 143.

\bibitem{GLM2} M. Gattobigio, A. Liguori and M. Mintchev, J. Math. Phys.
{\bf 40} (1999) 2949.

\bibitem{S} E. K. Sklyanin, J. Phys. A: Math. Gen. {\bf 21} (1988) 2375.

\bibitem{ZZ} A. B. Zamolodchikov and A. B. Zamolodchikov, Ann. Phys.
{\bf 120} (1979) 253.

\bibitem{F} L. D. Faddeev, Soviet Scientific Reviews Sect. C {\bf 1}
(1980) 107.

\bibitem{Fo} A. S. Fokas, Physica D {\bf 35} (1989) 167.

\bibitem{FM} L. Freidel and J.M. Maillet, Phys. Lett.
{\bf B262} (1991) 278.

\bibitem{MRSZ} M. Mintchev, E. Ragoucy, P. Sorba and Ph. Zaugg, J. Phys.
{\bf A32} (1999) 5885.

\bibitem{MW} S. Murakami and M. Wadati, J. Phys. {\bf A29} (1996) 7903.


\end{thebibliography}
\end{document}